\documentclass[12pt,letterpaper]{article}
\usepackage{graphicx}
\usepackage{amsmath,amssymb}
\usepackage{array,multirow}
\usepackage{booktabs}
\newcommand{\ra}[1]{\renewcommand{\arraystretch}{#1}}
\usepackage[margin=1in]{geometry}
\usepackage{adjustbox}
\usepackage{lineno}
\usepackage{soul}
\usepackage{color}

\title{Haze Evolution in Temperate Exoplanet Atmospheres Through Surface Energy Measurements}
\author{Xinting Yu*$^{,1}$, Chao He$^2$, Xi Zhang$^1$, Sarah M. H\"orst$^2$, \\
Austin H. Dymont$^3$, Patricia McGuiggan$^4$,  Julianne I. Moses$^5$, \\
Nikole K. Lewis$^6$, Jonathan J. Fortney$^7$, Peter Gao$^7$, \\
Eliza M.-R. Kempton$^8$, Sarah E. Moran$^2$, Caroline V. Morley$^9$, \\
Diana Powell$^7$, Jeff A. Valenti$^{10}$, V\'eronique Vuitton$^{11}$}
\date{%
    \small $^1$Department of Earth and Planetary Sciences, University of California Santa Cruz, 1156 High Street, Santa Cruz, California 95064, USA\\%
    $^2$Department of Earth and Planetary Sciences, Johns Hopkins University, 3400 N. Charles Street, Baltimore, Maryland 21218, USA\\%
    $^3$Department of Physics, University of California Santa Cruz, 1156 High Street, Santa Cruz, California 95064, USA\\%
    $^4$Department of Materials Science and Engineering, Johns Hopkins University, 3400 N. Charles Street, Baltimore, Maryland 21218, USA\\%
    $^5$Space Science Institute, Boulder, CO 80301, USA\\%
    $^6$Department of Astronomy and Carl Sagan Institute, Cornell University, 122 Sciences Drive, Ithaca, NY 14853, USA\\%
    $^7$Department of Astronomy and Astrophysics, University of California Santa Cruz, 1156 High Street, Santa Cruz, California 95064, USA\\%
    $^8$Department of Astronomy, University of Maryland, College Park, MD 20742, USA\\%
    $^9$Department of Astronomy, The University of Texas at Austin, Austin, TX 78712, USA\\%
    $^{10}$Space Telescope Science Institute, Baltimore, MD 21218, USA\\%
    $^{11}$University Grenoble Alpes, CNRS, CNES, IPAG, F-38000 Grenoble, France\\[2ex]%
    \today}
\begin{document}

\begin{titlepage}
\maketitle
\centering{*Email: xintingyu@ucsc.edu}\\
\centering{Accepted in Nature Astronomy}

\end{titlepage}

\begin{abstract}
Photochemical hazes are important opacity sources in temperate exoplanet atmospheres, hindering current observations from characterizing exoplanet atmospheric compositions. The haziness of an atmosphere is determined by the balance between haze production and removal. However, the material-dependent removal physics of the haze particles is currently unknown under exoplanetary conditions. Here we provide experimentally-measured surface energies for a grid of temperate exoplanet hazes to characterize haze removal in exoplanetary atmospheres. We found large variations of surface energies for hazes produced under different energy sources, atmospheric compositions, and temperatures. The surface energies of the hazes were found to be the lowest around 400 K for the cold plasma samples, leading to the lowest removal rates. We show a suggestive correlation between haze surface energy and atmospheric haziness with planetary equilibrium temperature. We hypothesize that habitable zone exoplanets could be less hazy, as they would possess high-surface-energy hazes which can be removed efficiently.
\end{abstract}

\section*{Introduction}

Photochemical hazes are significant components in planetary and exoplanetary atmospheres \cite{2017Natur.551..352Z, 2018NatAs...2..303H}. They could significantly impact the radiation budget of planetary atmospheres, affecting planetary climates \cite{2016AsBio..16..873A}. Recent work also suggests that hazes are the main particulate opacity sources in temperate exoplanetary atmospheres with equilibrium temperature (T$_{eq}$) less than 1000 K \cite{2020NatAs...4..951G}. Due to the existence of refractory aerosols, most transmission spectra obtained of the atmospheres of temperate Neptune-class and sub-Neptune exoplanets (T$_{eq}<$ 1000 K) are found to have diminished spectral amplitudes or even flat spectra in the current observable wavelengths (e.g., \cite{2014Natur.505...66K, 2014ApJ...794..155K, 2014Natur.505...69K, 2020AJ....159...57L}). On Earth, organic-rich atmospheric aerosols are produced in large amounts \cite{2000RvGeo..38..267J}; however, the terrestrial atmosphere is relatively clear in wavelengths and altitudes that can be probed by transmission spectroscopy for exoplanets. This is because the produced organic aerosols on Earth can be efficiently removed \cite{2020AGUA....100128B}. Thus, planetary haziness is likely a strong function of not only the production but also the removal of the haze particles. Knowing the removal efficiency of haze particles on habitable exoplanets is also important as the hazes transported to the surface of the planet may provide the initial organic matter to start extraterrestrial life \cite{2006PNAS..10318035T}. Diverse production rates of laboratory-made photochemical hazes have been observed under various atmospheric conditions \cite{2018NatAs...2..303H, 2018ApJ...856L...3H, 2018AJ....156...38H, 2020NatAs...4..986H, 2020PSJ.....1...51H}; however, the removal rates of the hazes remain unknown.

Atmospheric aerosols are mainly removed by dry and wet deposition \cite{2006book.conf.....S}. Dry deposition includes the processes of transporting aerosols directly to the surface or the deep atmosphere without the aid of precipitation, while wet deposition happens when aerosols are removed by cloud droplets. Both removal processes are material dependent. On Earth, many organic aerosols are hygroscopic and are easily removed \cite{2005ACP.....5.1053K}. In order to understand the removal physics of exoplanet haze particles and constrain the haziness of exoplanet atmospheres, here we assess one important material property that governs the removal of the hazes, the surface energy. The surface energy is defined as the energy needed to separate two solid surfaces per two-unit area, and it is an intrinsic material property arising from intermolecular interactions between contacting molecules \cite{2011book..122..432H}. The surface energy can help reveal many important properties of a material, including its bulk chemical structure, optical properties, cohesiveness, and wetting properties. The latter two respectively govern the dry and wet removal of haze particles. Here we measured the surface energies of a matrix of laboratory-produced exoplanet haze analogs via a simple measurement, the sessile drop contact angle method. Since the measurement requires only a few molecular layers of materials \cite{2011book..122..432H, 2005JG122.2610Y}, we are able to acquire surface energies for 18 exoplanet haze analog samples produced under a grid of atmospheric conditions (300--600 K, 100$\times$ to 10,000$\times$ metallicity), regardless of their production rates. Photochemical hazes, different from condensation clouds, are composed of thousands of chemical compounds \cite{2019ECS.....3...39H, 2020PSJ.....1...17M, 2020vuittonxxx} and are thus difficult to characterize. However, their bulk material properties such as the surface energy can be easily obtained through laboratory experiments and be incorporated into atmospheric microphysics models and observations.

\section*{Results}

\begin{figure}[!h]
\centering
\includegraphics[width=\textwidth]{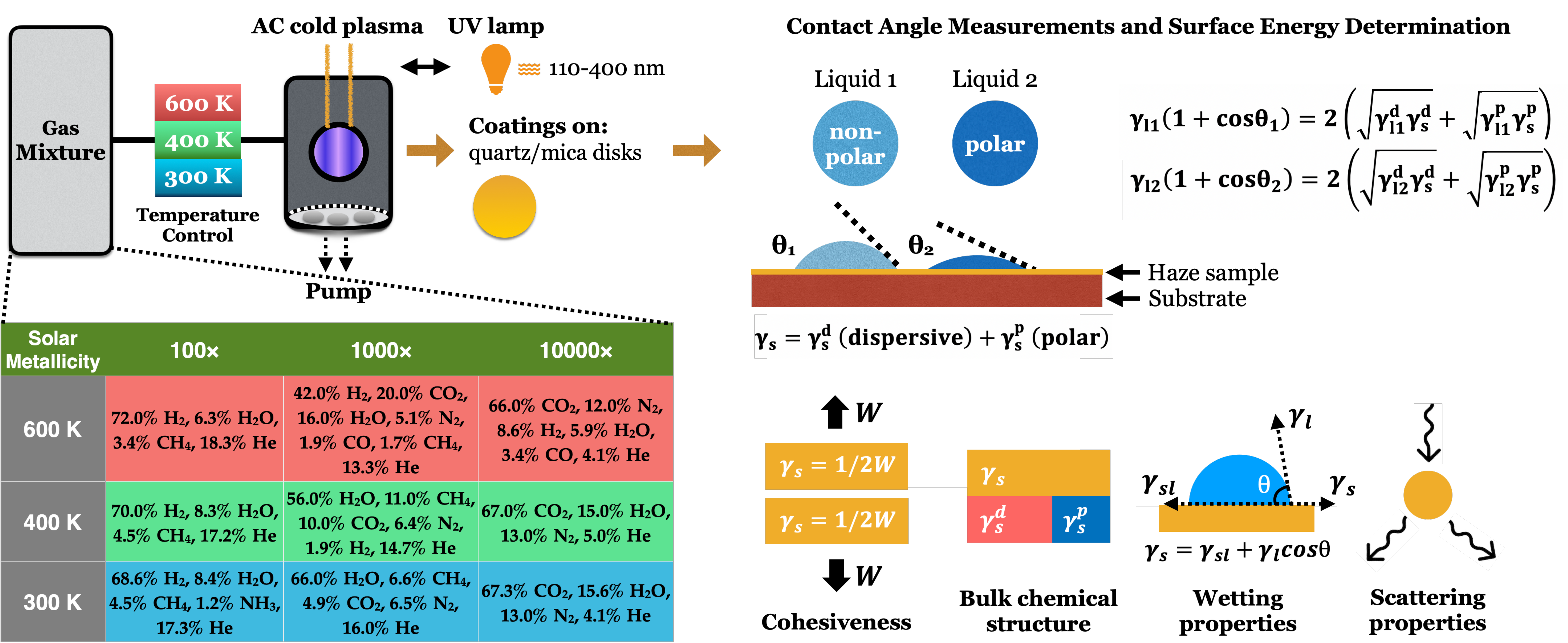}
\caption{\textbf{A schematic of the experimental setup for exoplanet haze production and surface energy measurement.} The initial gas mixtures were determined based on equilibrium composition for different temperatures (300, 400, 600 K) and solar metallicities (100$\times$, 1000$\times$, 10,000$\times$), where only the gas species with abundances above 1\% were considered for experimental manageability. For each experiment, one type of energy source was used, either an AC glow discharge cold plasma or a deuterium UV lamp. After 72 hours of reaction, solid haze samples were produced and deposited on the substrates. We used the sessile drop contact angle method to determine the surface energies of the haze samples. By measuring the contact angles formed between a haze-coated surface and one non-polar liquid and one polar liquid respectively, the total surface energies of the haze sample ($\gamma_s$) and its partitioning components (dispersive, $\gamma_s^d$ and polar, $\gamma_s^p$) can be calculated based on the known surface tensions of the test liquids ($\gamma_l$) and their partitioning components (dispersive, $\gamma_l^d$ and polar, $\gamma_l^p$) using the Owens-Wendt-Rabel-Kaelble (OWRK) method \cite{1969xxx..12..315C}. The total surface energy gives the total cohesiveness of a haze sample (which is half of the work of cohesion, W, the energy needed to separate two identical materials per unit area), and the partitioning components can inform us of the bulk composition of the sample. With the known surface tensions of other liquids, we can determine the wetting properties (contact angle $\theta$) between the haze sample and different clouds condensates in exoplanet atmospheres using the Young-Dupr\'e Equation ($\gamma_s=\gamma_{sl}+\gamma_l cos\theta$), where $\gamma_{sl}$ is the interfacial tension between the solid and the liquid. The surface energy can also be used to derive the refractive index of the sample and thus determine the scattering properties of haze particles in exoplanet atmospheres.}
\label{fig:setup}
\end{figure}

The haze analog samples were produced using the Planetary Haze Research (PHAZER) chamber \cite{2017ApJ...841L..31H} with temperatures and compositions of the initial gas mixtures summarized in Figure \ref{fig:setup}. The initial gas mixtures reflect the major atmospheric gas phase constituents (abundance $>$ 1\%, including H$_2$, He, H$_2$O, CH$_4$, N$_2$, CO$_2$, CO, and NH$_3$) at 1 mbar calculated for a range of solar metallicities (100$\times$, 1000$\times$, 10,000$\times$) and temperatures (300 K, 400 K, 600 K) based on a chemical equilibrium model \cite{2013ApJ...777...34M}. We used one of two energy sources to initiate chemistry for the gas mixture, a cold plasma generated by an alternating current (AC) glow discharge or a UV lamp with output continuum UV radiation from 110--400 nm, which leads to the production of solid haze particles that are deposited on substrates at the bottom of the chamber (see Methods). Each energy source produces 9 exoplanet haze analog samples, leading to a total of 18 samples, hereafter respectively referred as the cold plasma samples and the UV samples.

We use the sessile drop technique to measure the static contact angles between two test liquids (one polar, water, and one non-polar, diiodomethane) and the haze-coated surfaces (Figure \ref{fig:setup}). The measured contact angles are used to determine the surface energies of the haze samples (see Methods).

\subsection*{Total surface energies of the hazes--dry deposition}

\begin{figure}[!h]
\centering
\includegraphics[width=\textwidth]{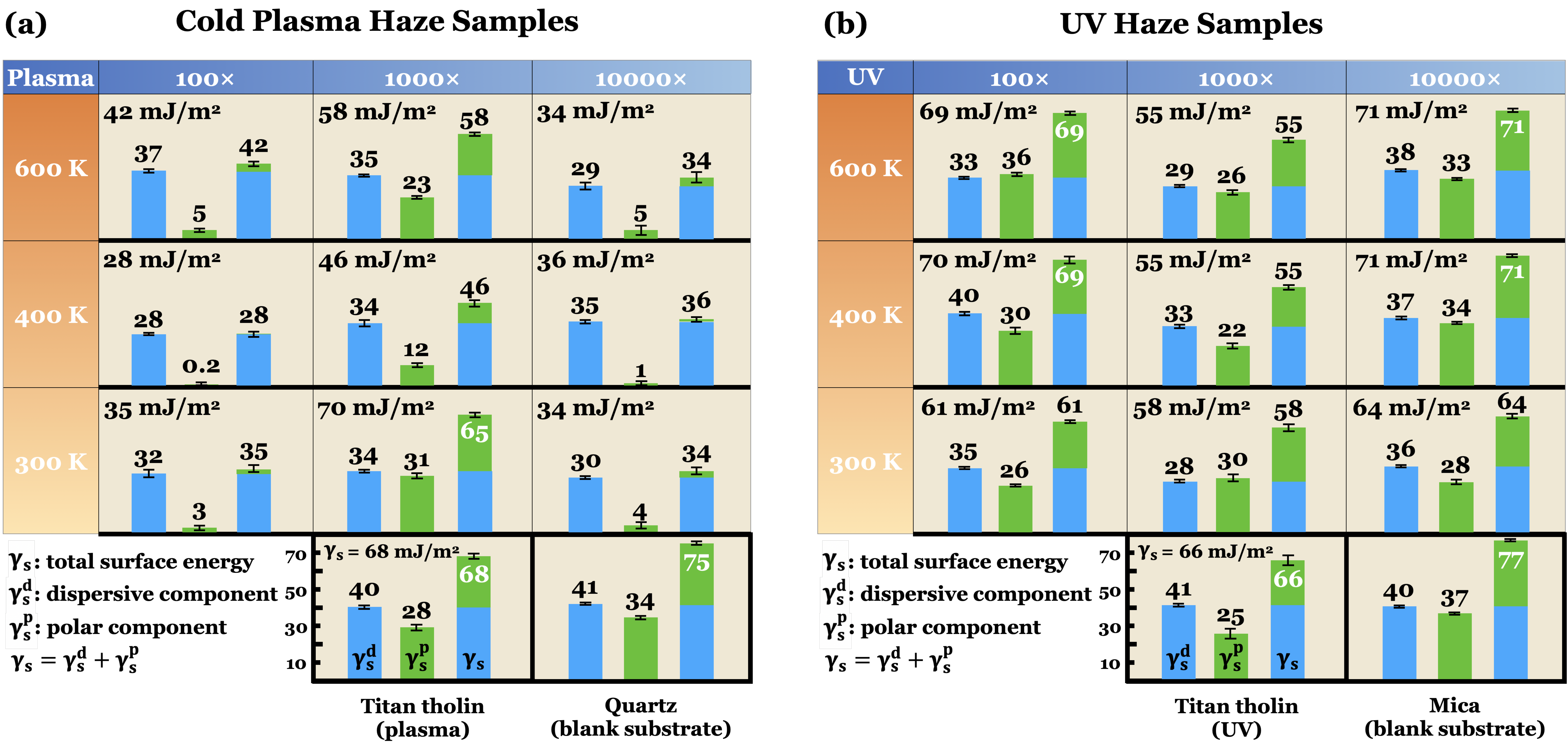}
\caption{\textbf{Summary of the derived surface energies for the cold plasma and UV exoplanet haze samples.} The surface energy values are derived using the OWRK two-liquid method, with unit of mJ/m$^2$. The blue, green, and stacked columns give the calculated mean values of the two partitioning components ($\gamma_s^d$ for the dispersive component, blue columns, and $\gamma_s^p$ for the polar component, green columns) and the total surface energies ($\gamma_s$, last column of each inset). The total surface energy for each sample are also labelled on the top left of each cell. All cells have the same scales as the Titan haze example (0--80 mJ/m$^2$). The surface energy of cold plasma and UV Titan tholin is adapted from \cite{2020apj.....1...51H}. The tabular results with standard deviations can be found in Supplementary Table 3 for the cold plasma samples and Supplementary Table 4 for the UV samples. The error bars of the total and the partitioning surface energy components represent 1-$\sigma$ s.d. uncertainties.}
\label{fig:summary}
\end{figure}

We summarize the derived total surface energies of the haze samples in Figure \ref{fig:summary}. The total surface energy of a material can indicate its cohesiveness. Titan haze analogs, produced with the same experimental setup, made with 95\% N$_2$/5\% CH$_4$ at 100 K, are measured to be highly cohesive (60--70 mJ/m$^2$, \cite{2020apj.....1...51H}) for both energy sources. In contrast, the total surface energies of the exoplanet hazes vary with energy sources.The difference in the total surface energies between the Titan haze analogs and the exoplanet haze analogs may be due to different haze formation pathways resulting from differences in the initial gas mixtures. The cold plasma samples have a wider range of surface energies (25--70 mJ/m$^2$, Figure \ref{fig:summary}(a)), while the surface energies of the UV samples have a smaller range and are all relatively cohesive (55--70 mJ/m$^2$, Figure \ref{fig:summary}(b)). At the same temperatures, for the cold plasma haze samples, the H$_2$O-dominated atmospheres (1000$\times$ metallicity) produce hazes with higher surface energy compared to ones produced under H$_2$- and CO$_2$-dominated atmospheres (100$\times$ and 10,000$\times$ metallicities). At the same metallicities, the cold plasma samples tend to have the lowest total surface energies at 400 K compared to either higher (600 K) or lower (300 K) temperatures, while the UV samples made under different reacting temperatures have similar total surface energies. 

The exoplanet haze samples are measured to have sizes between 20--180 nm (\cite{2018ApJ...856L...3H, 2018AJ....156...38H}). If the actual haze particles are similar in size to the laboratory-produced haze analogs, these particles would belong to the size range with the lowest dry removal rates \cite{1978AtmEn..12.2055S, 2020emerson}. These haze particles have to grow into larger particles ($> 1\ \mu m$) to be removed by gravitational settling. Interparticle forces between particles would enhance particle collisional rates, and this enhancement is more pronounced for small particles in thin upper atmospheres \cite{1981SurSc.106..529M}. Given the same roughness and stiffness, high-surface-energy hazes have higher interparticle forces than low-surface-energy hazes and would collide more frequently to grow into larger sizes \cite{2020montixxx}. Once the particles are deposited on the surface, high-surface-energy hazes would be harder to resuspend up into the atmosphere because they will have larger adhesion with the surface. For exoplanets without surfaces, high-surface-energy hazes, which more efficiently grow into larger particles, will sediment faster into the deep atmosphere where they are thermally decomposed. Thus dry deposition rates would be higher for high-surface-energy hazes compared to low-surface-energy hazes.

\subsection*{Surface energy partitioning--bulk chemical structure}

Partitioning of surface energy (Figure \ref{fig:summary}) can reveal the bulk chemical structures of the haze samples. The dispersive component is indicative of the strength of non-polar interactions. The polar component includes the polar interactions (dipole-dipole and H-bonding) and indicates the polarity of a material. We notice that the dispersive components of all the haze samples vary within a relatively small range (from 28--39 mJ/m$^2$), and the differences between the samples are mainly caused by their diverse polar components (from 0.2--36 mJ/m$^2$).

The polar components of the haze samples can be affected by the atmospheric composition and reacting temperature. For the cold plasma samples, they have a diverse range of polarities, from very polar (300 K, 1000$\times$, $\gamma_s^p=31$ mJ/m$^2$) to almost completely non-polar (400 K, 100$\times$, $\gamma_s^p=0.2$ mJ/m$^2$). Under the same temperatures, the H$_2$-rich and CO$_2$-rich atmospheres produce hazes that are less polar compared to the H$_2$O-rich atmospheres. Under the same metallicities, the 400 K samples are the least polar compared to hazes made with higher (600 K) or lower (300 K) temperatures. Even though the nine gas compositions are different for each experiment, a few pairs have the same gas species with slightly different abundances, such as the 600 K-100$\times$ and 400 K-100$\times$ cases, the 400 K-1000$\times$ and 300 K-1000$\times$ cases, and the 400 K-10,000$\times$ and 300 K-10,000$\times$ cases. Nevertheless, the 400 K cases always have the lowest polar components. All the UV samples have relatively high polarity, with a polar component of at least 20 mJ/m$^2$, which seems to be less affected by the reacting gas compositions and temperatures. The energy input can also affect the chemical structures of haze samples \cite{2018Icar..301..136H}. For the same gas mixture, we notice that the cold plasma samples generally have lower polar components compared to the UV samples.

The variations of the surface energy partitioning pattern among the samples are caused by the chemical structures of the solids, which are determined by the chemical processes happening in the experiments. All three factors in the experiments--energy source, atmospheric composition, and reacting temperature--will affect the chemistry. The polarities of the samples are related to how the polar oxygen- and nitrogen-containing species are incorporated into the solid samples. Both the cold plasma and the UV samples contain significant amounts of polar species as demonstrated by the solid phase elemental compositions (\cite{2020PSJ.....1...17M}). However, the polar components are not only determined by the bulk elemental compositions, but also by the specific bonding environments of the molecular structure. Though the UV lamp is unable to directly dissociate the triple bonds in N$_2$ and CO, previous works show that the products of these species are incorporated into the solid (e.g., \cite{2012AsBio..12..315T}). We believe that the UV lamp, through some unknown mechanisms, could be better at incorporating nitrogen  (from N$_2$ or NH$_3$) and oxygen (from H$_2$O, CO, or CO$_2$) into the solid sample and forming structures with higher polarity indices compared to the cold plasma samples. For example, the UV samples could have more high polarity structures such as R-OH, while the cold plasma samples could have more oxygen with low polarity structures such as R-O-R or R=O. This hypothesis can be validated with further structural characterization such as infrared (IR) or nuclear magnetic resonance (NMR) spectroscopy.

The UV samples are less affected by the reacting gas compositions and temperatures and are generally more polar than the cold plasma samples. Given that the commonality of all the nine gas mixtures is the existence of water vapor, one possible explanation is that the lower-energy UV lamp could be better at incorporating oxygen through water photolysis to form high polarity structures in the solids such as R-OH, while cold plasma could incorporate oxygen to form lower polarity structures such as R-O-R or R=O.

\subsection*{Haze-cloud interactions--wet deposition}

Haze particles could also be removed by condensable species in exoplanet atmospheres through wet deposition. Wet deposition occurs when the haze particles collide with a cloud droplet (impact scavenging) or become cloud condensation nuclei (CCN) for the cloud droplet (nucleation scavenging). Impact scavenging is the least efficient for particles with sizes between 50--1000 nm \cite{1978AtmEn..12.2055S}, as our hazes are \cite{2018ApJ...856L...3H, 2018AJ....156...38H}, and it typically occurs below cloud decks. Thus, the main wet deposition pathway for the haze particles is nucleation scavenging, where the haze particles are incorporated as CCN and are subsequently removed by rainout of condensation clouds. Nucleation scavenging is also the main way to remove aerosols on Earth (e.g., \cite{2005ACP.....5.1053K}). Thus we must investigate whether the hazes can be efficient nucleation seeds for condensation clouds to assess their wet removal rates.

Cloud nucleation for soluble and insoluble materials are respectively described by \cite{1936xxx..122..432H} and \cite{1962book..122..432H}. However, laboratory-made hazes are complex mixtures that include some portions of soluble molecules and large insoluble nuclei \cite{2020PSJ.....1...17M}. For partially soluble materials, we have to consider both their solubilities and surface properties to determine whether condensation clouds can efficiently nucleate on them \cite{1997JAerS..28..239G}. \cite{2002JGRD..107.4787R} found that a partially insoluble material can be a good source of CCN if 1) its solubility is over 10 mg/mL, or 2) its solubility is lower than 10 mg/mL but the material can be completely wetted by the condensate (or a contact angle $\theta=0^\circ$). 

\begin{figure}[!h]
\centering
\includegraphics[width=\textwidth]{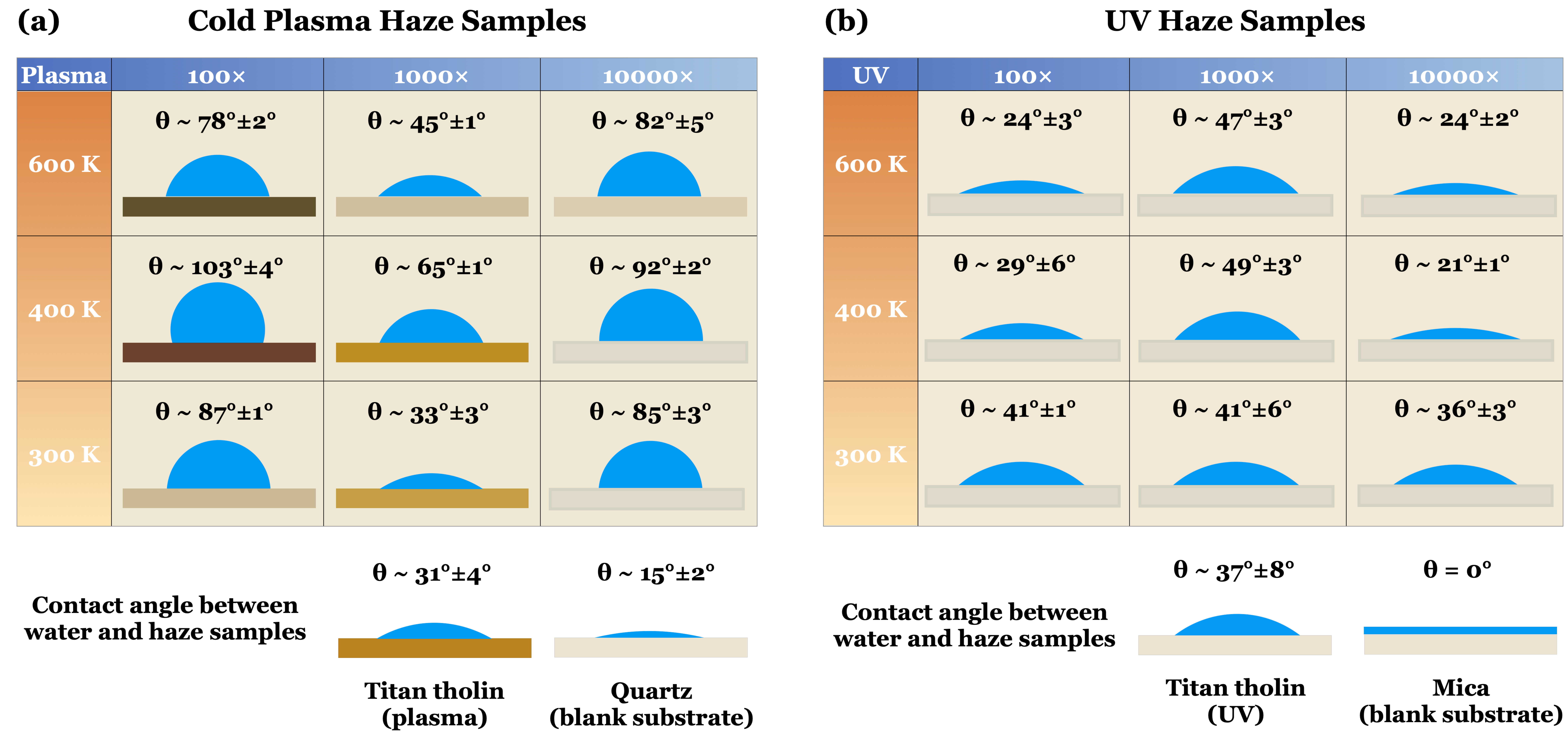}
\caption{\textbf{Summary of the measured mean contact angles between water and the haze samples using circle fitting results.} Colors of the blank quartz substrate, the exoplanet haze samples, and the Titan haze samples are adapted from \cite{2018ApJ...856L...3H}. The contact angle values between water and the Titan tholin samples are adapted from \cite{2020apj.....1...51H}. The UV samples and the mica substrate are denoted the same color as the blank quartz substrate since the production rate of the UV samples is too low to reveal visible colors. The error ranges of the contact angles represent 1-$\sigma$ s.d. measurement uncertainties from the fitting algorithm.}
\label{fig:water}
\end{figure}

Here we first determine the nucleation efficiency of the haze particles for liquid water clouds, as water is the key cloud condensate in the atmospheres of habitable zone exoplanets. We measured the solubilities for the 400 K, 1000$\times$ and the 300 K, 1000$\times$ cold plasma samples and performed qualitative solubility measurements for the rest of the exoplanet haze samples (see Methods). We found none of the samples have solubilities in water over 10 mg/mL. Thus, we have to examine the contact angles between the haze samples and liquid water to determine whether the hazes are good CCNs for liquid water clouds. We directly measured the contact angles between water and the haze samples (Figure \ref{fig:water}). Because water is a polar-dominated liquid (see Table \ref{table:liquids}), solids with low surface energies and especially low polar components will have high contact angles with liquid water. The hazes produced in H$_2$- and CO$_2$-dominated atmospheres have lower surface energies and lower polar components compared to the ones produced in H$_2$O-rich atmospheres, so their contact angles with water are higher. Similarly, the hazes produced at 400 K have higher contact angles with water compared to the ones produced at 300 K and 600 K.

According to \cite{2002JGRD..107.4787R}, a non-zero contact angle would lead to inefficient nucleation for materials below the solubility threshold. However, a higher than zero contact angle threshold could be adopted because of the natural existence of surface roughness on solid materials, which can lower the contact angle \cite{1936IdnE..122..432H}. All the laboratory haze samples are measured to be quite smooth \cite{2018ApJ...856L...3H, 2018AJ....156...38H} though the actual haze particles might be rougher. Because contact angle generally decreases with increasing surface roughness when $\theta\lessapprox$ 45$^\circ$ \cite{2008JCIS..325..472H}, here we adopt a generous threshold of 45$^\circ$, assuming rougher haze particles with contact angles less than 45$^\circ$ could become completely wettable. Two plasma haze samples (1000$\times$, 300 K and 600 K) and all the UV haze samples have small contact angles with water ($\theta\lessapprox$ 45$^\circ$), and thus they are all considered good CCNs for water clouds. Meanwhile, for larger contact angles, the addition of surface roughness would increase the contact angle. Smooth surfaces typically do not have $\theta>120^\circ$ \cite{1999Nishino..215..732L}. However, with surface roughness, a large contact angle can be formed, making a material ``super-hydrophobic" ($\theta>150^\circ$). The rest of the cold plasma exoplanet haze samples all have relatively large contact angles with water ($\theta>65^\circ$) and may potentially become super-hydrophobic aerosols in exoplanet atmospheres, unable to allow nucleation scavenging and water cloud growth.

\begin{table}
\caption{\textbf{Test liquids and their corresponding surface tensions and surface tension components.} All numbers have units of mN/m. $\gamma^{tot}$ is the total surface tension. $\gamma^{d}$ and $\gamma^{p}$ are respectively the dispersive and polar components used for the OWKR two-liquid method.}
 \label{table:liquids}
 \centering
 \ra{1.1}
 \begin{tabular}{c  c r c c}\toprule
 Liquid & Total surface tension && \multicolumn{2}{c}{OWRK components} \\
 \cmidrule{2-2} \cmidrule{4-5}
 & $\gamma_l^{tot}$ && $\gamma_l^d$ & $\gamma_l^p$\\
\midrule
Water & 72.8 && 21.8 & 51.0\\
Diiodomethane & 50.8 && 50.8 & 0\\
\bottomrule
 \end{tabular}
 \end{table}
 
Condensable species other than liquid water could also form clouds in exoplanet atmospheres, such as water ice, sulfuric acid (H$_2$SO$_4$), potassium chloride (KCl), zinc sulfide (ZnS), sodium sulfide (Na$_2$S), manganese sulfide (MnS), etc. \cite{2013ApJ...775...33M, 2018ApJ...863..165G, 2020RAA....20...99Z}. We do not know the solubilities of the hazes in these exotic condensates, so here we just assess the contact angle criteria to determine whether the hazes are good CCNs for these species, similar to previous works \cite{2020NatAs...4..951G}. With the measured surface energies of the haze samples, we can use the wetting theory to estimate the contact angles without direct measurements (see Methods). For water ice, H$_2$SO$_4$, and KCl, which condense in cooler atmospheres (T $\lessapprox$ 800 K), since their surface energies are around the same levels as liquid water (see Table \ref{table:exoplanet_liquids}), the contact angles formed between the hazes and these species should follow the same trend as liquid water, i.e., hazes with high surface energies form small contact angles and can be efficient CCN, and vice versa for hazes with low surface energies. For the cloud condensates in hotter atmospheres (T $\gtrapprox$ 800 K), including ZnS, Na$_2$S, MnS, chromium (Cr), forsterite (Mg$_2$SiO$_4$), iron (Fe), and aluminum oxide (Al$_2$O$_3$), because their surface tensions are all much larger ($\gamma_l>400$ mN/m, see Table \ref{table:exoplanet_liquids}) than even the haze sample with the highest surface energy, the resulting contact angle is at least 80$^\circ$. This means that the hazes should not be good CCN for these species. 

\begin{table}
\caption{\textbf{Possible condensates in exoplanet atmospheres, and their surface tension (for liquids)/surface energy (for solids) expressions.} The temperature ranges give where the surface tension/energy equations are applicable. The last column gives the calculated contact angle ranges between the condensates and the haze samples using Equation \ref{eq:cos}.}
 \label{table:exoplanet_liquids}
 \centering
 \ra{1.1}
\begin{adjustbox}{width=1.1\textwidth,center=\textwidth}

\begin{tabular}{c c c c c}\toprule
 Species & Temperature & Surface tension (liquid, mN/m) or & Reference & Contact angle \\
   &  range (K) & surface energy (solid, mJ/m$^2$) &  &  range ($^\circ$)\\
\midrule
H$_2$O (solid) & 273.15 & 106 & \cite{1974JCIS...46..185K} & 45--75\\
H$_2$SO$_4$ (liquid) & 283.15--925&  82.651(1 $-$ T/925)$\rm{^{1.2222}}$ & \cite{2009bookyaw....261C} & 0--58\\
KCl (liquid) & 700--1423.15 &  187.885(1 $-$ T/2600)$\rm{^{1.2227}}$ & \cite{2009bookyaw....261C} & 0--77\\
ZnS (solid) & n/a & 860 & \cite{2020NatAs...4..951G} & 85--88\\
Na$_2$S (solid) & n/a &1033 & \cite{2020NatAs...4..951G} & 86--88 \\
MnS (solid) & n/a & 2326 &  \cite{2020NatAs...4..951G} & 88--89 \\
Cr (liquid) & 1920--2218 & 3059.065(1 $-$ T/8560.93)$\rm{^{1.9841}}$ &  \cite{2009bookyaw....261C} & 88--89 \\
Mg$_2$SiO$_4$ (solid) & n/a & 436 & \cite{2020NatAs...4..951G}  & 81--86 \\
Fe (liquid) & 1811--2450 & 2707.417(1 $-$ T/9340)$\rm{^{1.6921}}$ &  \cite{2009bookyaw....261C} & 88--89  \\
Al$_2$O$_3$ (liquid) & 2323--2373 & 1148.443(1 $-$ T/6975)$\rm{^{1.2222}}$ &  \cite{2009bookyaw....261C} & 84--88   \\
\bottomrule
 \end{tabular}
  \end{adjustbox}
 \end{table}

Overall, for low-surface-energy hazes, their contact angles with the cloud condensates tend to be large, which make them less likely to be efficient CCN to be removed through wet deposition, and vice versa for high-surface-energy hazes. 

\subsection*{Refractive indices of the hazes}

\begin{figure}[!h]
\centering
\includegraphics[width=\textwidth]{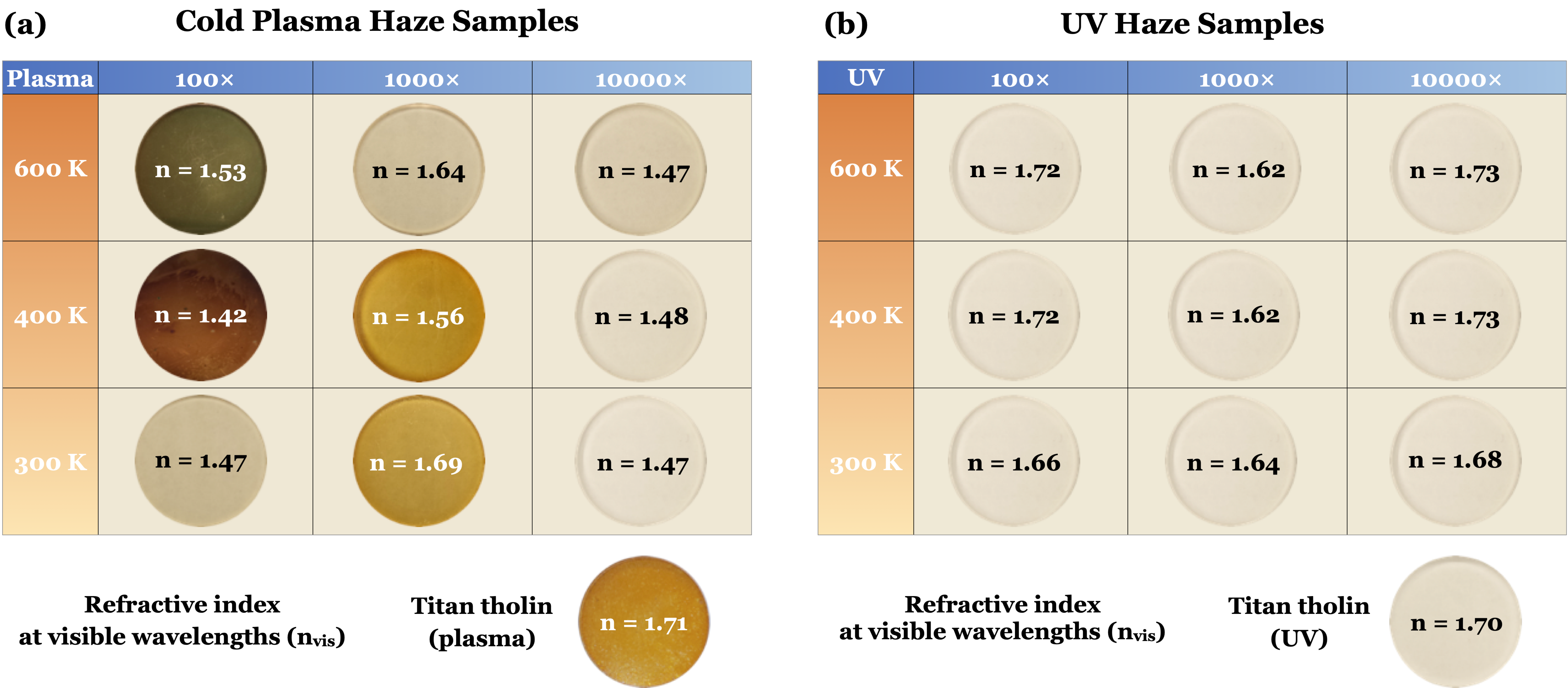}
\caption{\textbf{Summary of derived mean refractive indices at visible wavelengths (n$_{vis}$, see Methods) for the haze samples.} Pictures of the blank quartz substrate, the exoplanet haze samples, and the Titan haze samples are adapted from \cite{2018ApJ...856L...3H}. The refractive indices are derived using the simplified Lifshitz theory described in the Methods section. The error ranges of n$_{vis}$ represent 1-$\sigma$ s.d. uncertainties calculated through propagation of error.}
\label{fig:optical}
\end{figure}

Additionally, we can use the surface energies of the haze samples to derive the real part of the refractive indices of the hazes (n$_{vis}$, see Methods). The results are shown in Figure \ref{fig:optical}. Overall, the exoplanet hazes have a wide range of visible refractive indices, from 1.42--1.73. Because $n_{vis}$ increases with the total surface energy, the cold plasma samples have a wider range of $n_{vis}$ (from 1.42--1.72) while the UV samples have relatively high $n_{vis}$, from 1.62--1.73. The derived refractive indices would be useful for assessing the scattering properties of exoplanet hazes.

\section*{Discussion}

\begin{figure}[!h]
\centering
\includegraphics[width=\textwidth]{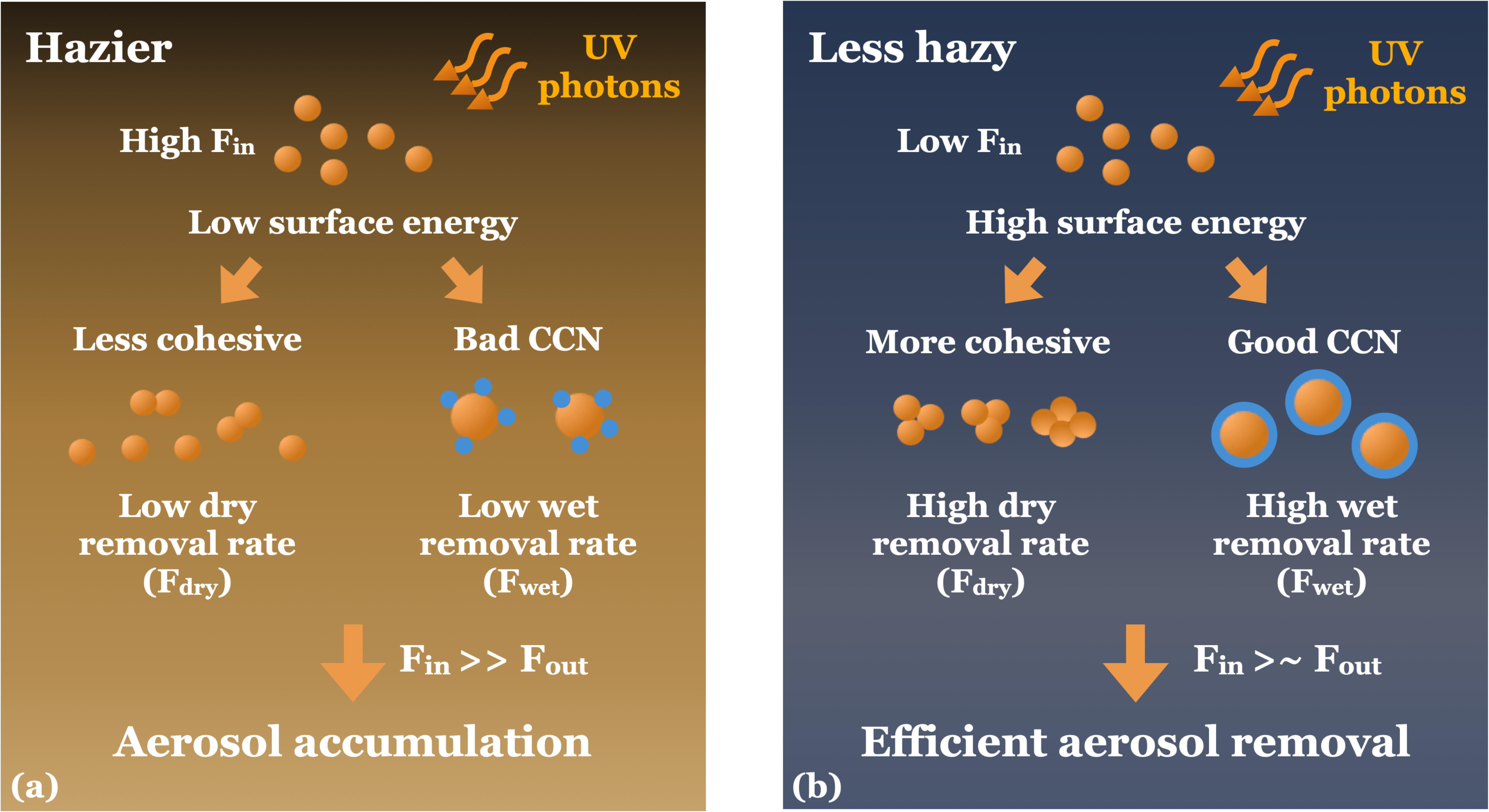}
\caption{\textbf{Haze production and removal schematics for low surface energy and high surface energy hazes.} The production rate and the removal rate of the hazes are denoted respectively as F$_{in}$ and F$_{out}$. The hazes either have low surface energies and are thus lyophobic (or hydrophobic for water), or have high surface energies and are thus lyophilic (or hydrophilic for water). (a) Low surface energy hazes are less cohesive and cannot coagulate efficiently to form larger particles to be removed through dry deposition, and they are usually not good CCN as well, thus they will not be removed efficiently by wet deposition. Thus the hydrophobic hazes will accumulate in planetary atmospheres after they are produced and make the atmospheres hazier over time. (b) The high surface energy hazes are more cohesive, and it is easier for them to coagulate into larger particles and be removed through gravitational settling (dry deposition). They are also good CCN for clouds and can be removed efficiently through nucleation scavenging (the main mechanism for wet deposition). Since the high surface energy haze particles can be removed more efficiently after they are produced, they will not accumulate as much in planetary atmospheres compared to the low surface energy hazes, leading to atmospheres that are less hazy.}
\label{fig:scheme}
\end{figure}

The haziness of an atmosphere is affected by both the production and removal of aerosols. Since photochemistry is constantly producing haze particles, if the formed particles are removed slowly, the atmosphere will end up being very hazy. If instead the hazes can be removed efficiently, the atmospheres would be less hazy. Overall, low-surface-energy hazes are less cohesive, have larger contact angles with cloud condensates, and are likely bad CCNs, making them harder to be removed by both dry and wet deposition (Figure \ref{fig:scheme}a). In this case, assuming other atmospheric transport properties such as vertical and horizontal mixing are similar and haze production rates are similar, the low-surface-energy hazes would accumulate, leading to hazier atmospheres. For high-surface-energy hazes, as they are more cohesive and have smaller contact angles with cloud condensates, they have higher removal rates (Figure \ref{fig:scheme}b). Thus, exoplanet atmospheres with high-surface-energy hazes could become less hazy. Condensable species are rare in some intermediate-temperature exoplanet atmospheres (300-600 K, \cite{2020RAA....20...99Z}), making wet deposition less likely. For these atmospheres, dry removal rates would still be higher for high-surface-energy hazes than the low-surface-energy hazes, but with wet removal rates being zero, the atmospheres would be hazier than the ones with condensable species. Note that atmospheric dynamics also affect the removal of the hazes, especially for wet deposition. For example, Titan's hazes are found to have high surface energy \cite{2020apj.....1...51H}; however, precipitation on Titan is infrequent and is concentrated mostly in the summer poles, leading to thinner hazes in the summer hemisphere compared to the winter hemisphere \cite{2010Icar..208..850R}.

\begin{figure}[!h]
\centering
\includegraphics[width=\textwidth]{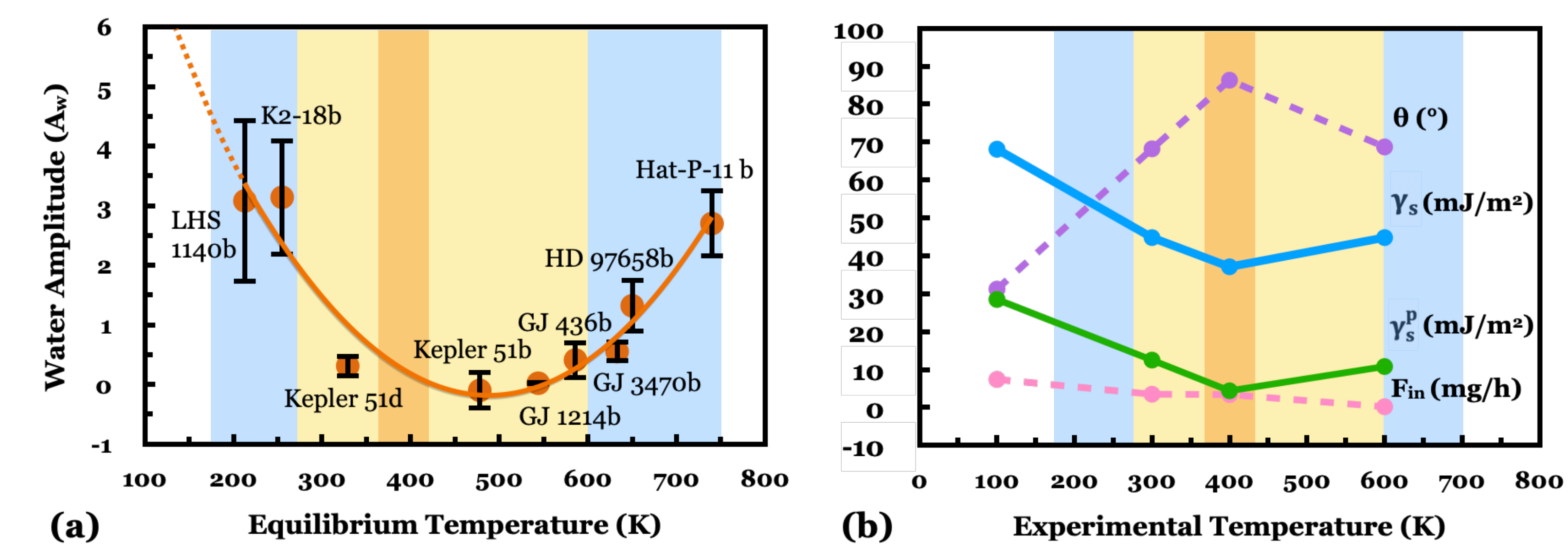}
\caption{\textbf{Exoplanet atmosphere haziness and properties of exoplanet haze samples as a function of temperature.} (a) Relationship between the mean exoplanet water absorption amplitude feature (A$_w$) and planetary equilibrium temperature (T$_{eq}$) for nine cool exoplanets with T$_{eq}<$ 800 K, assuming albedo $A=0.3$ \cite{2014Natur.505...66K, 2014Natur.505...69K, 2020AJ....159...57L, 2014Natur.513..526F, 2019NatAs...3..813B, 2019ApJ...887L..14B, 2020AJ....159..239G, 2021AJ....161...44E}. A$_w$ has a unit of scale height H, assuming a mean molecular weight of 2.3 amu. The fitted orange curve shows a best-fit second-order polynomial fit to the data with R$^2=0.91$. The error ranges of A$_w$ represent 1-$\sigma$ s.d. observation uncertainties. (b) Haze properties averaged over different metallicities (100$\times$, 1000$\times$, 10,000$\times$), including total surface energy ($\gamma_s$, blue solid line) in mJ/m$^2$, its polar component ($\gamma_s^p$, green solid line) in mJ/m$^2$, contact angle with water ($\theta$, purple dashed line) in $^\circ$, and production rates (F$_{in}$, pink dashed line) in mg/h \cite{2018NatAs...2..303H}. Only polar components of haze samples are shown because they dominantly determine the interactions between the hazes and water, a highly polar liquid. The yellow shading marks relatively hazy exoplanets between 270--600 K. Few condensable species exist at these temperatures to enable wet removal of haze particles \cite{2020RAA....20...99Z}. As wet removal is the dominant haze removal mechanism \cite{2005ACP.....5.1053K}, this leads to relatively hazy atmospheres. The orange shading marks the haziest observed exoplanets in panel (a) and the haze properties at this temperature in panel (b). Because this temperature includes both the lowest-surface-energy hazes and lacks condensable species, both dry and wet removal rates are the lowest, leading to the haziest atmospheres. The blue shading on the left marks the less hazy exoplanets (175--270 K, \cite{2011ApJ...736L..25K}), which includes the habitable zone where liquid water can condense in panel (a) and high-surface-energy hazes in panel (b). The blue shading on the right indicates less hazy hotter exoplanets where KCl condenses in panel (a) and high-surface-energy hazes in panel (b). In both blue shaded regions, hazes can be removed efficiently through both dry and wet removal.}
\label{fig:AH}
\end{figure}

In our experiment, we find that the surface energies of the hazes vary with a range of experimental conditions, including energy sources, gas compositions, and reacting temperatures. The UV haze samples generally have high surface energies, while the cold plasma samples have a range of surface energies with varying gas compositions and reacting temperature. A few cold plasma samples have very low surface energies and are even hydrophobic. These results have a few implications for the haziness of temperate exoplanets. The UV lamp we used here provided relatively low power and low energy density radiation (mainly lower-energy FUV/NUV radiation), while the cold plasma source generates energetic electrons, ions, and free radicals, simulating more energetic processes in planetary atmospheres. Currently, many exoplanets with atmospheric observations are orbiting closely to their host stars and experience relatively high energy environments \cite{2016ApJ...820...89F}. The chemical processes induced by this high energy environment may thus be better simulated by our cold plasma source. Therefore, here we assume that haze produced in these exoplanet atmospheres to be closer analogs to the cold plasma samples. In our experiment, we found that the 400 K samples are consistently less polar and have lower surface energies compared to both higher and lower temperature samples. Here we compiled the trend (Figure \ref{fig:AH}a) between equilibrium temperature and exoplanet haziness in terms of water absorption amplitude for nine Neptune-class and sub-Neptune exoplanets (with T$_{eq}$ less than 800 K) with transmission spectra measured by the G141 grism of Wide Field Camera 3 (WFC3) on the Hubble Space Telescope (HST) \cite{2014Natur.505...66K, 2014Natur.505...69K, 2020AJ....159...57L, 2014Natur.513..526F, 2019NatAs...3..813B, 2019ApJ...887L..14B, 2020AJ....159..239G, 2021AJ....161...44E}. A minimum of water amplitude is found around 400--550 K (the highest haziness), which corresponds with our experimental results that hazes produced at 400 K have relatively higher production rates (F$_{in}$, \cite{2018NatAs...2..303H}) and lower removal rates (F$_{out}$) because of their lower surface energies compared to the 300 K and 600 K samples (Figure \ref{fig:AH}b). Exoplanets that lack condensable species outside the habitable zone (300--600 K) are also likely to be hazy because of the lack of haze wet removal. We also include the surface energy data for the 100 K atmospheres with Titan hazes (made with 95\% N$_2$ and 5\% CH$_4$ \cite{2020apj.....1...51H}). The surface energy trend in Figure \ref{fig:AH}b suggests that we may be able to see additional less hazy habitable-zone exoplanets such as K2-18b and LHS 1140b \cite{2019ApJ...887L..14B, 2021AJ....161...44E}, where high surface energy hazes are produced and water condensation would promote wet removal. For exoplanets with higher temperatures ($\sim$600--750 K) with condensable KCl, their atmospheres could also be less hazy, as the high-surface-energy hazes can be efficiently removed by KCl clouds. Future searches for complex organic haze features in the infrared spectra of temperate exoplanetary atmospheres could validate our hypothesis. More laboratory experiments would better inform haze formation and evolution in exoplanet atmospheres.

\section*{Methods}
\subsection*{Material Preparation}

Exoplanet haze analog samples were produced with two different energy sources, a cold plasma \cite{2017ApJ...841L..31H, 2018ApJ...856L...3H} and a UV lamp \cite{2018AJ....156...38H, 2019ECS.....3...39H}, using the Planetary HAZE Research (PHAZER) experimental system at Johns Hopkins University. The initial gas compositions of the simulated exoplanets are summarized in Figure \ref{fig:setup}. They are calculated from the chemical equilibrium models of \cite{2013ApJ...777...34M} for exoplanets with a range of equilibrium temperatures (300, 400, 600 K) and solar metallicities (100$\times$, 1000$\times$, 10,000$\times$). AC glow discharge is a cold plasma energy source, which is used to simulate energetic processes in planetary and exoplanetary upper atmospheres \cite{2012chemi..195..792C}, and it can break all the molecular bonds of the gas molecules used in the experiments (\cite{2012chemi..195..792C}). The hydrogen UV lamp (HHeLM-L, Resonance LTD.) produces continuum UV irradiation from 110 and 400 nm which is typically used to simulate stellar far-UV (FUV, 91.2--170 nm) and near-UV (NUV, 170--320 nm) radiation for photochemistry, and it can directly dissociate single and double-bonded gas molecules (H$_2$, CH$_4$, H$_2$O, NH$_3$, CO$_2$), but not the triple-bonded N$_2$ and CO \cite{2006PNAS..10318035T, 2012AsBio..12..315T, 2014Icar..236..146S, 2018Icar..301..136H}. The UV lamp does indirectly dissociate triple-bonded molecules through secondary photochemical processes (e.g., \cite{2012AsBio..12..315T}). The energy density of the cold plasma is also higher than the UV lamp (170 W/m$^2$ versus 36 W/m$^2$, \cite{2019ECS.....3...39H}).

Gas mixtures that are listed in Figure \ref{fig:setup} for each experiment (excluding water) are mixed overnight in the mixing cylinder. Water vapor is supplied through vapor maintained by a cold bath of dry ice/methanol/water (see \cite{2018ApJ...856L...3H, 2018AJ....156...38H}). The prepared gas mixtures and water vapor were continuously flowed through a 15 m stainless-steel heating coil to reach the desired temperature. They then flowed through the reaction chamber with a flow rate of 10 standard cubic centimeters per minute (sccm) and were exposed to the energy source for around 3 s. The gases flowed continuously through the chamber for 72 hr and solid samples are deposited on the chamber walls and the substrates of choice.

The plasma samples are deposited on high-quality fused quartz disks (Ted Pella Inc., \cite{2018ApJ...856L...3H}) and the UV tholin samples were deposited on cleaved mica disks (Ted Pella Inc., \cite{2018AJ....156...38H}) on the bottom of the chamber. The resulting tholin films are relatively smooth for both energy sources. The root-mean-square (RMS) roughness over an area of 1 $\mu$m $\times$ 1 $\mu$m is 2--5 nm for the plasma tholin sample \cite{2018ApJ...856L...3H} and is 1--3 nm for the UV tholin sample \cite{2018AJ....156...38H}. The substrates were collected and stored in an oxygen and moisture free ($<$0.1 ppm O$_2$, $<$0.1 ppm H$_2$O) nitrogen glove box prior to measurements.

\subsection*{Contact angle measurements}
 
We used the sessile drop method to measure the contact angles between two test liquids and the coated exoplanet tholin surfaces. The test liquid was dispensed through a pipette to form a sessile droplet and was then gently placed onto the haze-coated surface. The contact angle between the film and the test liquid was recorded by a Ram\'e-Hart goniometer in ambient air (temperature 19--20 $^\circ$C, relative humidity 50--60\%). The test liquids used in our experiment include a polar liquid, HPLC-grade water (Fisher Chemical), and a non-polar liquid, diiodomethane (ReagentPlus\textsuperscript \textregistered, Sigma-Aldrich, 99\%). The surface tension of the test liquids and the corresponding components are listed in Table 1. 

We performed the measurements for each test liquid on the coated haze analog films two to six times on different areas of the film. An image for the droplet was recorded by the goniometer and the static contact angle was measured by using the ImageJ software \cite{2012naturemethods..122..432H} with the contact angle plugin. In ImageJ, we manually fit the drop profile by first choosing two points to define the baseline and then five points to define the drop profile. The contact angle plugin can fit the drop profile with either circle or ellipse approximation. The circle approximation is commonly used for small droplets or droplets with small contact angle. The ellipse approximation can more realistically fit larger droplets and droplets with larger contact angles, where the flattening of the drop profile due to gravity is more pronounced.. The drop size is controlled to be $<$2 $\mu$L to avoid the drop flattening due to gravity \cite{2010Lang...26H}.
 
\subsection*{Surface energy derivation methods}
The contact angle data obtained for each liquid can then be used to derive the surface energy of the tested material. \cite{2020apj.....1...51H} used five different methods to derive surface energy of Titan haze analogs and found the Owens-Wendt-Rabel-Kaelble (OWRK) two-liquid method to be the simplest and most accurate method for surface energy derivation for Titan hazes. This method is also the most widely-used method for surface energy determination \cite{2016booklz....261C}. Here we use the OWRK method to derive the surface energies of all the exoplanet haze samples \cite{1969xxx..12..315C, 1971Farbe...26H, 1970adhesion...77H}.

Assuming a liquid droplet and a solid surface are exposed to an inert atmosphere, the Young-Dupr\'e equation describes energy balance and the contact angle between the solid and the liquid phase:
\begin{equation}
W_{\rm{sl}}=\gamma_l^{\rm{tot}}(1+\rm{cos}\theta),
\label{eq:young-dupre}
\end{equation}
where W$_{\rm{sl}}$ is the work of adhesion between the solid and the liquid, $\gamma_{\rm{tot}}$ is the total surface tension of the liquid, and $\theta$ is the contact angle formed between the liquid and the solid.

In order to use Equation \ref{eq:young-dupre} to solve for the surface energy of the solid with the measured contact angle data, the OWRK method approximates the work of adhesion W$_{\rm{sl}}$ into the surface tension/energy components of the liquid and the solid:
\begin{equation}
W_{\rm{sl}}=2(\sqrt{\gamma_s^d\gamma_l^d}+\sqrt{\gamma_s^p\gamma_l^p}),
\label{eq:owrk}
\end{equation}
where $\gamma_s^d$ and $\gamma_s^p$ are the dispersive and polar components of the total surface energy of the solid, and $\gamma_l^d$ and $\gamma_l^p$ are the dispersive and polar components of the total surface tension of the liquid. The dispersive component includes the non-polar molecular interactions (London dispersive forces) and the polar component includes dipole-dipole and H-bonding interactions. The total surface tension of the liquid ($\gamma_l^{\rm{tot}}$) and surface energy of the solid ($\gamma_s^{\rm{tot}}$) can be expressed as:
\begin{equation}
\gamma_l^{\rm{tot}}=\gamma_l^d+\gamma_l^p,
\label{eq:liquid}
\end{equation}
\begin{equation}
\gamma_s^{\rm{tot}}=\gamma_s^d+\gamma_s^p.
\label{eq:solid}
\end{equation}
 
Combining Equation \ref{eq:young-dupre} and \ref{eq:owrk}, we can get:
\begin{equation}
\gamma_l^{\rm{tot}}(1+\rm{cos}\theta)=2(\sqrt{\gamma_s^d\gamma_{l}^d}+\sqrt{\gamma_s^p\gamma_{l}^p}),
\label{eq:contactangle}
\end{equation}
Thus, with the contact angle measurement of two liquids on the same solid surface, we can retrieve the dispersive and the polar component of the solid surface energy by solving two sets of Equation \ref{eq:contactangle}.

\subsection*{Solubility determination}
We determined the solubility (S) for the 400 K, 1000$\times$ and 300 K, 1000$\times$ cold plasma haze samples quantitatively with solid samples. We measured and transferred respectively 5.8 mg and 5.5 mg of the 400 K, 1000$\times$ and the 300 K, 1000$\times$ solid samples into two sample vials. We then added 0.5 mL of HPLC-grade water (Fisher Chemical) into each vial and sonicated both vials for approximately 1 hr. The sonicated mixtures were then filtered by 0.1 $\mu$m PTFE hydrophilic syringe filters (with weight $m_{initial}$) to collect the remaining solid. The syringe filters were dried overnight in a 50$^\circ$C oven to remove the excess water and the remaining solids were weighed within the syringe filter ($m_{final}$). The solubilities of the haze samples can then be determined:
\begin{equation}
S=\frac{m_{final}-m_{initial}}{0.5\  mL},
\end{equation}
which is 4.6 mg/mL and 5.0 mg/mL respectively for the 400 K, 1000$\times$ and the 300 K, 1000$\times$ cold plasma samples. Note that the initial solution we used was more concentrated (11.6 mg/mL and 11.0 mg/mL) than the solubility threshold (10 mg/mL) proposed by Raymond and Pandis (2002). However, large amount of materials were left for both haze samples, indicating that both haze samples are below the solubility threshold.

For the partially soluble haze samples, when a water droplet was placed on these samples, the contact angles of water would decrease with time. Supplementary Figure 1 plots the change in contact angle with time for water on each haze sample. For the two samples with measured high solubilities (300 K, 1000$\times$, 5.0 mg/mL and 400 K, 1000$\times$, 4.6 mg/mL), the decreases in contact angles are respectively 7.4$^\circ$ and 9.8$^\circ$. The only sample with higher contact angle decrease is the 600 K, 10,000$\times$ sample, 11.1 $^\circ$. However, it is unlikely that this sample will exceed the solubility threshold of 10 mg/mL, as the solubility of Titan tholin is 6.0 mg/mL and its contact angle with water decrease is 17.6$^\circ$ \cite{2020apj.....1...51H}. Thus, we consider all the haze samples to be below the solubility threshold of 10 mg/mL and we need to assess their contact angle with time to determine whether or not they are good CCNs for water clouds. Note that some samples have their contact angles with water increasing with time, which is likely due to the measurement error associated with the drop profile (fitting error, denoted as the error bar in Supplementary Figure 1 and manually profile choice error).

\subsection*{Refractive index retrieval}
The simplified Lifshitz theory \cite{1992lif....261C, 2011book..122..432H} established the link between the surface energy and the refractive index of a material at visible wavelengths (n$_{vis}$). With the derived the surface energies from the previous section, we can then use the Lifshitz theory of van der Waals forces to retrieve n$_{vis}$ of the hazes. The Lifshitz theory treats forces between two surfaces as continuous media and are derived with bulk properties of the materials, including their dielectric constants and refractive indices. The main terms of the resulting van der Waals force coefficient, the Hamaker constant for three media is:
\begin{equation}
A\approx\frac{3}{4}kT(\frac{\varepsilon_1-\varepsilon_3}{\varepsilon_1+\varepsilon_3})(\frac{\varepsilon_2-\varepsilon_3}{\varepsilon_2+\varepsilon_3})+\frac{3h}{4\pi}\int_{\nu_1}^{\infty}(\frac{\varepsilon_1(i\nu)-\varepsilon_3(i\nu)}{\varepsilon_1(i\nu)+\varepsilon_3(i\nu)})(\frac{\varepsilon_2(i\nu)-\varepsilon_3(i\nu)}{\varepsilon_2(i\nu)+\varepsilon_3(i\nu)})d\nu,
\label{eq:lifshitz}
\end{equation}
where where k is the Boltzmann constant, T is the temperature of the system, h is the Planck's constant, $\varepsilon_1$, $\varepsilon_2$, and $\varepsilon_3$ are the static dielectric constants of the three media, $\varepsilon_1(i\nu)$, $\varepsilon_2(i\nu)$, and $\varepsilon_3(i\nu)$ are the dielectric constants at imaginary frequencies, and $\nu_n=(2\pi kT/h)n$, where n is the quantum number of the relevant oscillation.

The dielectric constants can be approximated as a function of frequency using an interpolation formula proposed by \cite{1970biophy...26H}:
\begin{equation}
\varepsilon(i\nu)=1+\frac{\varepsilon_0-\varepsilon_{mw-ir}}{1+\nu/\nu_{mw}}+\frac{\varepsilon_{mw-ir}-\varepsilon_{vis}}{1+\nu^2/\nu_{ir}^2}+\frac{\varepsilon_{vis}-1}{1+\nu^2/\nu_{uv}^2},
\label{eq:dielectric}
\end{equation}
where $\nu_{mw}$, $\nu_{ir}$, and $\nu_{uv}$ are respectively the characterization absorption frequencies at the microwave ($<$10$^{12}$ s$^{-1}$), infrared ($\sim$10$^{14}$ s$^{-1}$), and ultraviolet (3$\times$10$^{15}$ s$^{-1}$). $\varepsilon_0$ is the static dielectric constant at zero frequency, $\varepsilon_{mw-ir}$ is the dielectric constant where microwave relaxation ends and the infrared relaxation begins, and $\varepsilon_{vis}$ is the dielectric constant at the visible wavelengths, where infrared relaxation ends and the UV relaxation begins. $\varepsilon_{vis}$ can be linked to the refractive index of the medium in the visible wavelength through the Maxwell relationship (n$_{vis}^2=\varepsilon_{vis}$, \cite{1865RSPT..155..459C}).

Equation \ref{eq:dielectric} can be simplified as the majority of the interactions originates from the electronic excitation in the UV frequency range \cite{2011book..122..432H}, thus the dielectric constants can be written as:
\begin{equation}
\varepsilon(i\nu)=1+\frac{n_{vis}^2-1}{1+\nu^2/\nu_{uv}^2}.
\label{eq:dielectric_simple}
\end{equation}

Integration of the second term of Equation \ref{eq:lifshitz} is now possible. Because $\nu_1$ is usually much smaller than the main adsorption frequency $\nu_{uv}$, the lower integration limit $\nu_1$ is replaced by 0, and integration of Equation \ref{eq:lifshitz} leads to a Hamaker constant of:
\begin{equation}
A\approx\frac{3}{4}kT(\frac{\varepsilon_1-\varepsilon_3}{\varepsilon_1+\varepsilon_3})(\frac{\varepsilon_2-\varepsilon_3}{\varepsilon_2+\varepsilon_3})+\frac{3h\nu_{uv}}{8\sqrt{2}}\frac{(n_1^2-n_3^2)(n_2^2-n_3^2)}{\sqrt{n_1^2+n_3^2}\sqrt{n_2^2+n_3^2}(\sqrt{n_1^2+n_3^2}+\sqrt{n_2^2+n_3^2})},
\label{eq:lifshitz_simple}
\end{equation}

For cohesion, where media 1 and 2 are identical, we can rewrite Equation \ref{eq:lifshitz_simple} into:
\begin{equation}
A\approx\frac{3}{4}kT{(\frac{\varepsilon_1-\varepsilon_3}{\varepsilon_1+\varepsilon_3})}^2+\frac{3h\nu_{uv}}{16\sqrt{2}}\frac{(n_1^2-n_3^2)^2}{(n_1^2+n_3^2)^{3/2}}.
\label{eq:lifshitz2}
\end{equation}
If media 3 is vacuum or inert air, we have $\varepsilon_3=1$ and $n_3=1$, thus Equation \ref{eq:lifshitz2} can further be simplified to:
\begin{equation}
A\approx\frac{3}{4}kT{(\frac{\varepsilon_1-1}{\varepsilon_1+1})}^2+\frac{3h\nu_{uv}}{16\sqrt{2}}\frac{(n_1^2-1)^2}{(n_1^2+1)^{3/2}}.
\label{eq:lifshitz3}
\end{equation}
Because the first term is only a few percent of the second term, we can approximate the Hamaker constant to be:
\begin{equation}
A\approx\frac{3h\nu_{uv}}{16\sqrt{2}}\frac{(n_1^2-1)^2}{(n_1^2+1)^{3/2}}.
\label{eq:lifshitz4}
\end{equation}

The Hamaker constant and the surface energy of a material are linked by the following equation:
 \begin{equation}
\gamma=\frac{A}{24\pi d_0^2},
\label{eq:hamaker_surface}
\end{equation}
where $d_0$ is equilibrium separation distance for two surfaces at contact, and is found to be around 0.157$\pm$0.009 nm for most materials \cite{2006bookvan....261C}. 

Using Equation \ref{eq:lifshitz4} and \ref{eq:hamaker_surface}, we can retrieve the refractive index of the tholin films. Note that the derived n$_{vis}$ using this method is only an approximation for a broad visible wavelength range instead of any specific wavelength. The theory is found to work well for most solids and liquids except for highly polar H-bonding liquids such as water (error within 10--20\%, \cite{2011book..122..432H}), as H-bonding forces are not considered in the Lifshitz theory. Here we assume the surface energies of the hazes do not have any H-bonding contributions, and use the total surface energy of the hazes to derive their refractive indices. We found the theory works well for deriving the refractive indices of Titan tholins (made with carbon, nitrogen, and hydrogen) using their total surface energies (E. Sciamma-O'Brien, private communications; C. He, private communications). It is possible that we overestimate n$_{vis}$ of the exoplanet haze samples by using their total surface energies, as in addition to nitrogen, carbon, and hydrogen, they are also oxygen-rich \cite{2020PSJ.....1...17M}, which may lead to a more significant H-bonding component in the total surface energies than the Titan tholins. 

\subsection*{Contact angle estimation}

We can use the Young-Dupr\'e equation to retrieve the contact angle between the solid and liquid, given their surface energy and surface tension, the Young-Dupr\'e equation (Equation \ref{eq:young-dupre}) can also be written as:
\begin{equation}
cos\theta=\frac{\gamma_s-\gamma_{sl}}{\gamma_l},
\label{eq:cos}
\end{equation}
where $\gamma_{sl}$ is the interfacial tension between the solid and the liquid phase, and it is assumed to be zero. When cos$\theta\geq1$, the liquid would completely wet the solid surface. While when cos$\theta\approx0$, the contact angle is very large ($\theta\approx90^\circ$). We estimated the contact angles between the measured surface energies of the haze samples and the known surface tension of the cloud condensates and results are summarized in Table \ref{table:exoplanet_liquids}.

\subsection*{Temperate exoplanets haziness trend}

We compiled recent transmission spectra data for temperature exoplanets with T$_{eq}<800$ K and calculated strengths of the water feature for each exoplanet to represent their cloudiness or haziness.  The water amplitude calculation was based on previous trend studies \cite{2016ApJ...817L..16S, 2017ApJ...847L..22F, 2017AJ....154..261C}. We selected nine temperate exoplanets with equilibrium temperature less than 800 K with transmission spectra taken by the near-infrared G141 grism of the WFC3 on the HST \cite{2014Natur.505...66K, 2014ApJ...794..155K, 2014Natur.505...69K, 2020AJ....159...57L, 2014Natur.513..526F, 2019NatAs...3..813B, 2019ApJ...887L..14B, 2020AJ....159..239G, 2021AJ....161...44E, 2019NatAs...3.1086T}. We choose the water feature centered at 1.4 $\mu$m and the baseline continuum at 1.25 $\mu$m to derive the water amplitude for each exoplanet. Following \cite{2016ApJ...817L..16S}, we computed the weighted average of the transit depths ($\sqrt{\Delta D}=R_p/R_*$) around 1.25 $\mu$m from 1.22--1.3 $\mu$m, and around 1.4 $\mu$m from 1.36--1.44 $\mu$m. We then converted the transit depths to the radius of the planet (R$_p$) at the particular wavelengths and then took the difference between R$_p$ at 1.25 $\mu$m and 1.4 $\mu$m to get the water amplitude strength. The water amplitude was finally normalized by the scale height of the exoplanet following \cite{2017AJ....154..261C}. This method is a simple way of directly estimating the water amplitude of exoplanet atmospheres with the current available observational data. Figure \ref{fig:AH} shows the relationship between the water amplitude features and planetary equilibrium temperature (calculated assuming albedo $A=0.3$ and a mean molecular weight of 2.3 amu).

\section*{Acknowledgments}
X.Y. is supported by the 51 Pegasi b Fellowship from the Heising-Simons Foundation. X.Z. is supported by NASA Solar System Workings Grant 80NSSC19K0791. C.H. and S.M.H. are supported by the NASA Exoplanets Research Program Grant NNX16AB45G. P.M. is supported by the 3M Nontenured Faculty Grant. S.E.M. is supported by NASA Earth and Space Science Fellowship grant 80NSSC18K1109.

\section*{Author contributions}
X.Y., C.H., X.Z., S.M.H., conceived the study. C.H. prepared the samples. X.Y. and C.H. performed the surface energy measurements. P. M. helped with interpretation for the surface energy measurements. X.Y. and A.H.D. performed the water amplitude calculations. X.Y., C.H., S.M.H., and J.I.M. discussed the chemistry for haze formation. X.Y. and X.Z. discussed the implication of the results to haze removal. X.Y. conducted the data analysis and prepared the manuscript. X.Y., C.H., X.Z., S.M.H., A.H.D., P.M., J.I.M., N.K.L., J.J.F., P.G., E.M.R.K., S.E.M., C.V.M., D.P., J.A.V., and V.V. contributed to discussing the results and editing the manuscript.

\section*{Competing interests}
The authors declare no competing interests.

\section*{Data availability}
The data used to create Figures \ref{fig:summary}--\ref{fig:optical} are available in the supplementary data tables. The original data used to create Figure \ref{fig:AH} can be found in the supplementary data tables and the source papers cited in the manuscript. The original sessile drop contact angle image files and the processed contact angle and surface energy data files can be found in the repository with the following DOI: https://doi.org/10.7291/D1BM2P.

\end{document}